\documentclass[pra,twocolumn,titlepage,nofootinbib,amsmath,showpacs]{revtex4}%
\usepackage{graphicx}%
\newcommand{\eq}[1]{(\ref{#1})}
\newcommand{\alphab}{\mbox{\boldmath$\alpha$}}
\newcommand{\Za}{Z\alpha}
\newcommand{\Zab}{(Z\alpha)}
\newcommand{\calO}[1]{{\cal O}\left( #1 \right)}
\newcommand{\someletter}{\varepsilon}

\begin{document}
\title{Relativistic recoil effects in a muonic atom within a Grotch-type approach: General approach}
\author{Savely~G.~Karshenboim}
\email{savely.karshenboim@mpq.mpg.de}
\affiliation{Max-Planck-Institut f\"ur Quantenoptik, Garching,
85748, Germany} \affiliation{Pulkovo Observatory, St.Petersburg,
196140, Russia}
\author{Vladimir G. Ivanov} \affiliation{Pulkovo
Observatory, St.Petersburg, 196140, Russia}
\author{Evgeny Yu. Korzinin}
\affiliation{D.~I. Mendeleev Institute for Metrology, St.Petersburg,
190005, Russia}

\begin{abstract}
Recently we calculated relativistic recoil corrections to the energy
levels of the low lying states in muonic hydrogen   induced
by electron vacuum polarization effects. The results were obtained by
Breit-type and Grotch-type calculations. The former were described in
our previous papers in detail, and here we present the latter.

The Grotch equation was originally developed for pure Coulomb
systems and allowed one to express the relativistic recoil correction to
order $(Z\alpha)^4m^2/M$ in terms of the relativistic non-recoil
contribution $(Z\alpha)^4m$. Certain attempts to adjust the method
to electronic vacuum polarization took place in the past, however,
the consideration was incomplete and the results were incorrect.

Here we present a Groth-type approach to the problem and in a series of papers
consider relativistic recoil effects in order
$\alpha(Z\alpha)^4m^2/M$ and $\alpha^2(Z\alpha)^4m^2/M$. That is the
first paper of the series and it presents a general approach, while
two other papers present results of calculations of the
$\alpha(Z\alpha)^4m^2/M$ and $\alpha^2(Z\alpha)^4m^2/M$
contributions in detail. In contrast to our previous calculation, we
address now a variety of states in muonic atoms with a certain range
of the nuclear charge $Z$.
\end{abstract}
\maketitle


\section{Introduction}

Spectroscopy of light muonic atoms was used for a while and provided
us with certain important data on the nuclear structure. It was
based on a study of the emission lines and had limited accuracy.
Recently, the first successful laser-spectroscopy measurement on
muonic hydrogen has  opened a new generation of experiments. The
experiment performed at PSI delivered the value of the Lamb shift in
muonic hydrogen and allowed to determine the proton charge radius
with unprecedented accuracy. Unexpectedly, that measurement has led to one of
the currently largest controversies in QED related experiments. A strong
discrepancy between the value of the proton charge radius obtained
from muonic hydrogen \cite{nature} and that in ordinary hydrogen
\cite{codata2010} is of about 5 standard deviations. Meantime, the latter
value is in perfect agreement with a recent electron-proton scattering
result \cite{mainz}.

That circumstance has renewed interest in spectroscopy of muonic
atoms. The low $l$ states and, mostly, the $1s$ and $2s$ states are
sensitive to the finite-nuclear-size effects and have been used for
a while to determine the charge radius for a broad range of nuclei
from hydrogen \cite{nature} to uranium \cite{zumbro}.

Higher-$l$ states are also of interest for more ``metrological''
measurements. In particular, the $3d_{5/2}{-}2p_{3/2}$ transition in
muonic $^{24}$Mg and $^{28}$Si was used in \cite{mumass} to
determine $m_\mu/m_e$. A similar measurement was also performed
in pionic atoms to
determine the pion mass. In such experiments one has to deal with
X-ray transitions and then there is a problem in calibration of
the X-ray standards. In \cite{pimass} the $5f{-}4g$ transition in
pionic nitrogen and the $6h{-}5f$ one in pionic neon were compared
with $5f{-}4g$ transitions in muonic oxygen.

Higher $l$ states can also be of interest due to antiprotonic helium
spectroscopy. At present, highly accurate data are available only for
a three-body system, which includes a nucleus, antiproton and
electron \cite{mep2:1,mep2:2}. While the antiproton in a circular or
a near circular state is rather immune against annihilation, the
electron ``protects'' the antiprotonic state from collision quenching.
Still, a possibility for a two-body antiprotonic helium ion has not
been given up and such a system may be of experimental interest in
the future.

In this situation a theoretical study of low-lying states of circular
states, such as $2p, 3d, 4g, 5f, 6h$ is of practical interest. Since
the muon mass is substantially higher than the electron mass, one
has to pay attention to recoil effects.

To find recoil contributions to energy levels of a hydrogenic atom
one can apply various approaches and, in particular, a Grotch-type
one.

A calculation of recoil corrections to order $m/M$ is possible in
hydrogenic atoms exactly \cite{shabaev} (see also \cite{yelkhovsky})
without any expansion in $Z\alpha$. The result consists of two
contributions, one is a result of one-photon exchange in an
effective Dirac equation, while the other takes into account
multi-photon exchanges.

It is the one-photon exchange that was first derived in
\cite{Gro67} without any expansion in $Z\alpha$. The Grotch equation
is an efficient way to derive from the one-photon-exchange term the
result which allows one to combine a few important features of theory of
the energy levels and to obtain a result which incorporates
\begin{itemize}
\item the leading nonrelativistic term (i.e. a result of the
Schr\"odinger-Coulomb problem) exactly in $m/M$;
\item the complete relativistic series for
infinitely heavy nucleus  (i.e. a result of the Dirac-Coulomb
problem) exactly in $(Z\alpha)$;
\item the leading relativistic recoil correction to energy in order
$(Z\alpha)^4m^2/M$.
\end{itemize}

On the other hand, the electronic vacuum polarization (eVP) effects
and, in particular, the Uehling potential, play a crucial role in
the theory of energy levels in muonic atoms. It is important to be able
to calculate relativistic and recoil corrections to them for a
variety of levels.

Recently, such a relativistic recoil contribution of order
$\alpha(Z\alpha)^4m^2/M$ was considered in various approaches for
low-lying states in light muonic atoms \cite{jentschura,borie11,pra}
(see also \cite{pach96,pach04,borie05} for earlier evaluations). Results on
$\alpha^2(Z\alpha)^4m^2/M$ can be found in \cite{a2Za4}.

Here, we rederive the Grotch equation for a pure Coulomb problem
and generalize it for a broad class of potentials. The generalized
approach allows one to find relativistic recoil eVP corrections in the
first and second order in $\alpha$, which are studied in subsequent
papers \cite{II,III}.

\section{One-photon exchange in two-body bound systems\label{s:ope}}

The Coulomb bound two-body systems have a binding energy of order
$(Z\alpha)^2m$, where $\alpha$ is the fine structure constant, $Z$
is the nuclear charge, $m$ is the mass of the orbiting particle,
i.e. the lighter one in the bound system. Throughout the paper
we apply relativistic units in which $\hbar=c=1$. These energy levels
have various corrections due to the relativistic, recoil and QED
effects and due to the nuclear structure.

The $(Z\alpha)^2m$ term can be found by many different methods,
while the methods to derive the corrections often depend on the
nature of those corrections. A certain class of the corrections can
be expressed in terms of the potentials and one can expect that for
their evaluation it is possible to adjust approaches used for
pure Coulomb calculations.

The potential corrections and, in particular, those presented by
the Uehling potential, are dominant QED effects for light and
medium-$Z$ muonic atoms. Here we develop an effective approach to
study relativistic recoil corrections in the first order in the
electronic vacuum polarization.

Electronic vacuum polarization (eVP) effects are responsible for the
Uehling potential, but even for the relativistic recoil contribution one
has to go somewhat beyond just the Uehling potential, just as for the
calculation of the $(Z\alpha)^4m^2/M$ term one has to go beyond a
pure Coulomb field. Here $M$ is the nuclear mass and appearance of
the $m/M$ ratio indicates that recoil effects are involved.

Throughout the paper we consider a point-like nucleus; however, in
many situations the finite-nuclear-size effects can be treated as
a small perturbation and, specifically, for low-$Z$ calculations and
for a high $l$ medium-$Z$ case. In any case, the finite nuclear size
affects the interaction between the muon and the nucleus; however,
the effect can be still described as a kind of potential and the
results obtained below can be in part adjusted for the extended
nuclei.

Relativistic recoil effects contribute to one-photon exchange as
well as to many-photon exchanges. The Coulomb and
Uehling potentials correspond to a dominant contribution in
one-photon exchange.

\begin{figure}[htpb]
\begin{center}
\resizebox{0.14\textwidth}{!} {\includegraphics{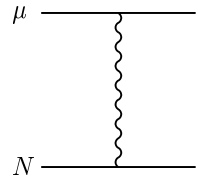}}
\end{center}
\caption{The leading one-photon-exchange diagram. It is responsible for
the contributions to orders $(Z\alpha)^2m$ and $(Z\alpha)^4m$.
\label{f:c1gamma}}
\end{figure}

\begin{figure}[htpb]
\begin{center}
 \resizebox{0.14\textwidth}{!}
{\includegraphics{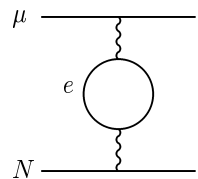}}
\end{center}
\caption{The one-photon-exchange diagram for the eVP contributions.  It
is responsible for the the Uehling-potential corrections to orders
$\alpha(Z\alpha)^2m$ and $\alpha(Z\alpha)^4m$. \label{f:u1gamma}}
\end{figure}

The one-photon contribution for the Coulomb case and Uehling term
are depicted in Figs.~\ref{f:c1gamma} and~\ref{f:u1gamma},
respectively. They are responsible for the entire nonrelativistic
contribution to orders $(Z\alpha)^2m$ and $\alpha(Z\alpha)^2m$,
respectively.

Those contributions can be described by a potential. They partly
include recoil effects in a sense, that one has to use the reduced
mass $m_R = m M/(m+M)$ in calculations. The result for the
Uehling correction can be achieved analytically in terms of
elementary functions \cite{uehl_cl,uehl_an}. The potential approach can
be also applied for a relativistic evaluation with the Dirac wave
functions. For the Uehling potential the energy with the Dirac wave
functions is known in closed analytic terms \cite{rel1,rel2}.

Indeed, as far as the wave functions for Schr\"odinger-Coulomb and
Dirac-Coulomb problems and the dispersion
presentation of the Uehling potential, such as
\begin{equation}\label{e:uehling}
V_U(r) = -\frac{\alpha\Zab}{\pi} \int_0^1
  dv \, \rho_e(v) \frac{e^{-\lambda r}}{r}
  \,,
\end{equation}
where
\begin{eqnarray}
\lambda&=&\frac{2m_e}{\sqrt{1-v^2}}
  \,,\nonumber\\
  \rho_e(v)&=&\frac{v^2(1-v^2/3)}{1-v^2}\,,
\end{eqnarray}
are well known,
a numerical calculation has never been a problem (see, e.g.,
\cite{pach96,jentschura,borie11}). Nevertheless, analytic
evaluations allow one to find various useful asymptotics
\cite{rel1,uehl_an,rel2}.

We note that the Uehling potential is smaller than the Coulomb
potential roughly by a factor of $\alpha/\pi$ in any kinematic area.
Similarly, we see that the eVP potential related to the second-order
correction possesses the same property --- it is smaller than
the Coulomb exchange in any kinematic area by a factor of
$(\alpha/\pi)^2$. Since general behavior of the eVP-induced
potentials is somewhat similar to the $(\alpha/\pi)V_C(r)$ and
$(\alpha/\pi)^2V_C(r)$, we can hope that whatever we use for a pure
Coulomb problem, it may be adjusted for eVP effects, including the
relativistic recoil.

\begin{figure}[htbp]
\begin{center}
  \resizebox{0.40\textwidth}{!} {
  \includegraphics{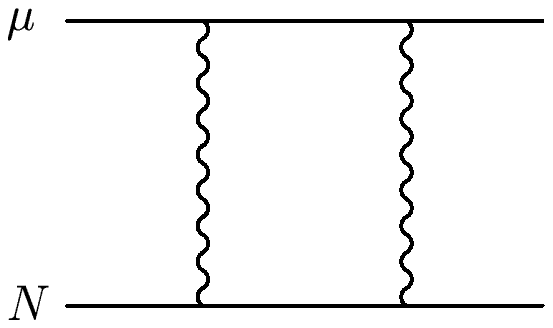} \resizebox{0.03\textwidth}{!} {\ } \includegraphics{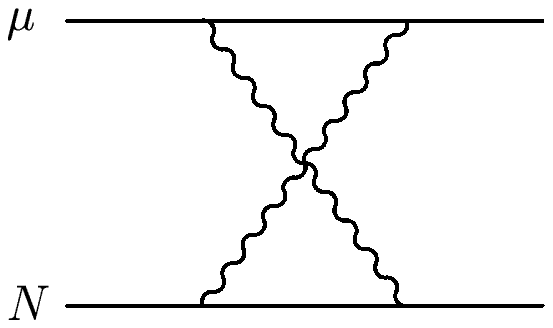}
  }
\end{center}
\caption{The leading two-photon-exchange diagrams.
In case of any practical calculations, there should be subtraction
terms due to the nonperturbative nature of the Coulomb exchange for the
bound state problem; meanwhile some one-photon-``reducible''
contributions can appear. Those are not shown here.
In certain gauges and, in particular, in the Feynman gauge the
two-photon-exchange term contributes to order $(Z\alpha)^4m^2/M$,
while in the Coulomb gauge it contributes only to order
$(Z\alpha)^5m^2/M$. \label{f:c2gamma}}
\end{figure}

\begin{figure}[htbp]
\begin{center}
  \resizebox{0.40\textwidth}{!} {
  \includegraphics{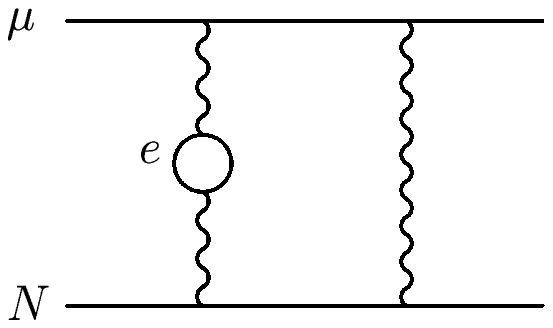} \resizebox{0.03\textwidth}{!} {\ } \includegraphics{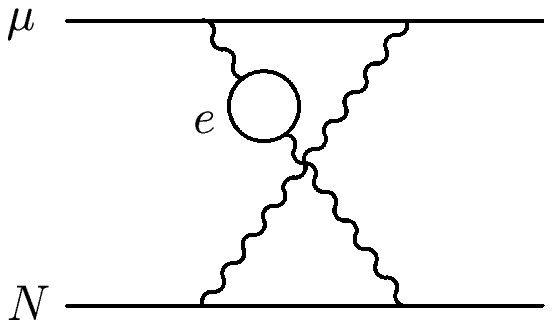}
  }
\end{center}
\caption{Two-photon-exchange diagrams for the eVP contribution.
Subtraction terms and reducible contributions are omitted. In
certain gauges the two-photon-exchange effects contribute to order
$\alpha(Z\alpha)^4m^2/M$. \label{f:u2gamma}}
\end{figure}

Meanwhile, neither a complete calculation of the one-photon exchange
(Figs.~\ref{f:c1gamma} and \ref{f:u1gamma}) can be identically
presented in terms of a potential, nor can the two-photon one exchange
(Figs.~\ref{f:c2gamma} and \ref{f:u2gamma}) be in general ignored.

The approach developed by Grotch and Yennie \cite{Gro67} allowed one to
resolve this problem for exchange by free photons
(Figs.~\ref{f:c1gamma} and \ref{f:c2gamma}) and here
\cite{pra} we generalize it,  following our
previous paper, for the case of the eVP
contributions.

At first, we have to address a question of a possibility to use a
certain relativistic equation with a kind of an effective potential
for a calculation of recoil effects.

The one-photon-exchange contribution can be evaluated with the help of
the photon propagator, which in the Coulomb gauge takes the form
\begin{eqnarray}\label{g:coul}
D_{00}^C&=&-\frac{1}{{\mathbf k^2}}\;, \nonumber\\
D_{i0}^C&=&0\;,\nonumber\\
D_{ij}^C&=&-\left(\delta_{ij}-\frac{k_ik_j}{{\mathbf
k^2}}\right)\;\frac{1}{k^2}\;,
\end{eqnarray}
where $k^2=k_0^2-{\mathbf k}^2$. Note, that only the static part of
$D_{00}$ produces a contribution in the non-recoil limit and thus is
responsible for an electrostatic potential. The other components of the
photon propagators in general depend on the choice of the gauge and
they are not directly related to $D_{00}$. For this reason the
complete one-photon contribution cannot in general be expressed in
terms of an electrostatic potential.

The one-photon contribution in the Coulomb gauge can be reduced for
the $m/M$ correction to its static approximation (i.e. neglecting
the $k_0$ dependence) and thus to several potential-like terms
because
\begin{itemize}
\item there is no $k_0$ dependence in $D_{00}$ and thus no
retardation effects are involved (if they were involved, that
still would be of reduced importance because they are proportional to
$k_0^2/{\mathbf k}^2$ and for atomic energy levels that would lead
to relativistic corrections proportional
to $(m/M)^2$, while here we are interested
in the $m/M$ correction only);
\item $D_{i0}=0$;
\item $D_{ij}$ involves lower components of the
spinor for the nucleus and thus the contribution is proportional to
at least $m/M$, which means that the retardation effects in the
$D_{ij}$ term are of order $(m/M)^2$ or higher and negligible.
\end{itemize}
In the next sections we apply the static approximation to the
one-photon exchange and develop an effective potential equation,
first for a pure Coulomb problem and next for a perturbed Coulomb
problem.

The remaining question is about two-photon-exchange contributions
for the $(Z\alpha)^4m^2/M$ correction. (In any effective Dirac
equation approach, and we follow such an approach since we are to
find a Grotch-type effective Dirac equation, it is assumed that
certain two-photon-exchange subtractions take place (see, e.g.,
\cite{ede} for detail).) This question was reviewed, e.g. in
\cite{pra}. The two-photon contributions are of at least order
$(Z\alpha)^5m^2/M$ in the Coulomb gauge because
\begin{itemize}
\item there is no $k_0$ dependence in $D_{00}$ and thus there is
no photon pole in two-photon exchange
with two $D_{00}$ components;
\item $D_{i0}=0$, and thus there is no contribution which involves
one $D_{i0}$ photon and one $D_{00}$ photon.
\end{itemize}
That is sufficient to avoid any potential $(Z\alpha)^4m^2/M$
contribution.

\section{Grotch equation and its solution for the Coulomb bound
systems\label{s:gc}}

Once we are limiting our consideration to one-photon-contribution in
a static approximation (i.e. at $k_0=0$), we can derive the Grotch
equation for the free one-photon exchange (Fig.~\ref{f:c1gamma}) in
order, after that, to generalize it step by step for a more general
case, including the eVP contributions. Our consideration closely
follows the original one by Grotch and Yennie \cite{Gro67}.

Here we give a brief reminder of the derivation of the Grotch
equation and its solution in order to describe every step which we
will need to adjust to eVP contributions.


The Grotch equation \cite{Gro67} is one of several effective Dirac equations
for a two-particle system. It is important to reproduce the two most
important features of any system of two fermions with an orbiting
particle much lighter than the nucleus. The electron in ordinary
hydrogen and the muon in muonic hydrogen are such particles. It is
useful to consider the orbiting particle within a full relativistic
consideration, while treating the nucleus in the leading
nonrelativistic approximation. As a result, we may derive an
equation, which correctly reproduces its limits both the
Schr\"odinger-Coulomb equation with the reduced mass and the
Dirac-Coulomb equation with the original mass of the muon (or
electron). Indeed, the equation is also supposed to take into
account certain relativistic recoil corrections. The uncertainty in
the calculation of the static one-photon contribution is of order
$(Z\alpha)^4(m/M)^2m$. The two-photon
contribution is of order $(Z\alpha)^5m^2/M$.

The desired equation is of the form of Dirac equation for a muon
\begin{equation}
 \left[\widehat{P}_n-\widehat{p}_N-m -
 \widetilde{V}_{1\gamma}\right]\Psi_n=0\,.
\end{equation}
where
$\widehat{A}=\gamma^\nu_{[\mu]} A_\nu$ and $P_n = (E_n, {\bf 0
)}$ (here $A^\nu$ is an arbitrary vector, $\nu$ is a relativistic 4-index,
while $\mu$ stands for a muon.)

This is an equation in the center-of-mass system. While the equation
is for the muon energy and wave function, the quantized energy $E_n$
is for the two-body system and we should subtract from the whole
4-momentum $P_n$ the nuclear 4-momentum $p_N=\left(\sqrt{M^2+{\bf
p}^2}, -{\bf p}\right)$, where ${\bf p}$ is the muon momentum.

To obtain a one-particle equation from a two-body one it was suggested
that one can present the two-body wave function $\Psi_{\mu N}$ in
terms of the free nuclear spinor and the muon wave function $\psi$
\begin{equation}\label{e:psi}
\Psi_{\mu N}={1 \choose -\frac{{\bf p}\cdot{\mbox{\boldmath
$\sigma$}}_N}{2M}} \psi\;.
\end{equation}
This suggestion is not just an approximation in a sense that one can
construct a perturbation theory and systematically take into account
all the corrections required for a certain level of accuracy. The
nuclear on-shell corrections are of relativistic nature for the
nucleus and thus they are of higher order in $m/M$ and $Z\alpha$
than the leading recoil effects we study. The off-shell corrections
can be found through many-photon exchange diagrams and a proper
choice of gauge can eliminate them in the leading recoil order.

The effective potential $ \widetilde{V}_{1\gamma}$ results from the
static part of the one-photon exchange averaged over the nuclear
part of the wave function in (\ref{e:psi}). In the momentum space we
find
\begin{eqnarray}\label{wv}
 \widetilde{V}_{1\gamma}({\bf q},{\bf p})&=&
 -i\gamma_{\mu}^0 \gamma_{N}^0(Z\alpha)
 \left(1, -\frac{{\mbox{\boldmath $\sigma$}_N}\cdot{\bf q}}{2M}\right)
 \nonumber\\
 &&\times
 \left[i\gamma_{N}^0\gamma_{\mu}^0D_{00}({\bf k})+i\gamma_{N}^i\gamma_{\mu}^jD_{ij}({\bf k})\right]
 {1 \choose -\frac{{\mbox{\boldmath $\sigma$}_N}\cdot{\bf p}}{2M}}
 \nonumber\\\nonumber\\
 &=&-\frac{Z\alpha}{{\bf k}^2}
 \Bigg\{1+\frac{1}{2M}\bigg[{\mbox{\boldmath $\alpha$}}_\mu\cdot({\bf p}+{\bf q})
 \nonumber\\
 &&\phantom{9}
 -\frac{({\mbox{\boldmath $\alpha$}}_\mu\cdot{\bf k})\ ({\bf k}\cdot({\bf p}+{\bf q}))}{{\bf k}^2}\bigg]
 \nonumber \\
 &&-\frac{1}{2M}[{\bf k}\times i{\mbox{\boldmath $\sigma$}}_N]\cdot{\mbox{\boldmath $\alpha$}}_\mu\nonumber\\
 &&+\calO{(Z\alpha)^4\left(\frac{m}{M}\right)^2m}\Bigg\}\,,
\end{eqnarray}
where ${\bf k} = {\bf p} - {\bf q}$. Here, the neglected term is not
$\calO{(Z\alpha)^4({m}/{M})^2m}$ by itself, but it represents an operator,
the matrix element of which over the atomic wave function is
$\calO{(Z\alpha)^4({m}/{M})^2m}$.

This effective potential includes a nuclear-spin-dependent term
which is responsible for the hyperfine splitting. It is of order
$(Z\alpha)^4m^2/M$. However, experimentally and theoretically the
hyperfine structure effects are well separated from the Lamb shift
effects. We consider this term as a perturbation and neglect the
hyperfine-interaction term (i.e. we average over the nuclear spin).

Once we average the results over the nuclear spin, i.e. over the
hyperfine structure, we note that all the remaining nuclear-spin
effects appear only in order $(m/M)^2$ (see, e.g.,
\cite{spin0,spin1,spin_nov,spin_r}) and thus this derivation,
started for the nuclear spin 1/2, is now valid for a nucleus with an
arbitrary spin.

That is the last crucial step to obtain the Grotch equation
\cite{Gro67} and we arrive at that in coordinate
space
\begin{eqnarray}\label{Dirac_C}
  &&\biggl(
    \alphab \cdot {\mathbf p}
    + \beta m
    + \frac{\mathbf{p}^2}{2M}
    + V_C
    + \frac{1}{2M} \left\{ \alphab \cdot {\mathbf p}, V_C
    \right\}\nonumber\\
    &~& +\frac{1}{4M}
      \left[
         \alphab \cdot {\mathbf p},
         [\mathbf{p}^2, W_C]
      \right]
  \biggr) \psi(r)
  =
  E \psi(r)
  \,.
\end{eqnarray}
where the operator $W_C$ appears due to taking into account the
$D_{ij}^C$ components of the photon propagator. It is essential
that it can be expressed in a certain way through $V_C$, which is
defined through $D_{00}^C$.
In particular, for free one-photon-exchange (Fig.~\ref{f:c1gamma})
and the relation between the
Coulomb gauge is of the form%
\begin{equation}\label{c_wc}
  W_C({\mathbf k}) = -\frac{2V_C({\mathbf k})}{{\mathbf k}^2}
  \,.
\end{equation}
For the case of the Coulomb gauge one finds in coordinate and
momentum space
\begin{eqnarray}
  V_C(r) &=& -\frac{\Za}{r}\,,\nonumber\\
  V_C({\mathbf k}) &=& -\frac{4\pi\Za}{{\mathbf k}^2}
   \,,
\end{eqnarray}
and
\begin{eqnarray}\label{wc}
  W_C(r) &=& -\Za r
  \,,\nonumber\\
  W_C({\mathbf k}) &=& \frac{8\pi\Za}{{\mathbf k}^4}
  \,.
\end{eqnarray}
While the leading part of $D_{00}^C$ in any gauge should produce the
Coulomb term $V_C$, the shape of the Hamiltonian in Eq.~(\ref{Dirac_C})
and a particular shape of $W_C$ depends on the gauge chosen.

The effective equation above can be solved in a closed analytic form
after applying a series of transformations \cite{Gro67}.
%
%
We start with rearranging the Hamiltonian
\begin{eqnarray}\label{Dirac_CH}
  H&=&\biggl(
    \alphab \cdot {\mathbf p}
    + \beta m
    + \frac{\mathbf{p}^2}{2M}
    + V_C
    + \frac{1}{2M} \left\{ \alphab \cdot {\mathbf p}, V_C
    \right\}\nonumber\\
    &~&~~~ +\frac{1}{4M}
      \left[
         \alphab \cdot {\mathbf p},
         [\mathbf{p}^2, W_C]
      \right]
  \biggr)
\end{eqnarray}
as following
\begin{equation}\label{Dirac2_C}
  H =
  H_0+\delta H
    + {\cal O} \left(  \Zab^4 \frac{m^3}{M^2}\right)
  \,,
\end{equation}
where
\begin{equation}\label{c:hoh1}
  H_0=H_1
  + \frac{H_1^2-m^2}{2M} + \frac{1}{4M} [H_1,[\mathbf{p}^2,W_C]]
  \,,
\end{equation}
\begin{equation}\label{c:h1}
  H_1 = \alphab \cdot {\mathbf p} + \beta m + V_C \frac{1-\beta m/M}{1-(m/M)^2}
  \,,
\end{equation}
and
\begin{equation}\label{c:dh}
  \delta H =
  -\left( \frac{V_C^2}{2M} + \frac{1}{4M} [V_C,[\mathbf{p}^2,W_C]] \right)
      \,.
\end{equation}

The correction, neglected  in (\ref{Dirac2_C}), is indeed an
operator; its matrix elements over bound states are of order ${\cal
O} \left(  \Zab^4\frac{m^3}{M^2} \right)$, which is explicitly shown
in (\ref{Dirac2_C}). In this sense Eq.~(\ref{Dirac2_C}) is not
correct as an operator identity, but it is sufficiently valid for
all matrix elements for the bound states.

We note that due to the relation between $V_C$ and $W_C$ (\ref{c_wc})
the last term vanishes for the Coulomb potential in the Coulomb
gauge
\begin{equation}\label{c:dh0}
  \delta H=0\;.
\end{equation}

To solve Eq.~(\ref{Dirac_C}) within the required accuracy is the
same as to solve equation
\begin{equation}
  H_0\psi_0=E_0 \psi_0
  \,,
\end{equation}
where $E=E_0$ and $\psi=\psi_0$ for the pure Coulomb case.

To deduce $E_0$ and $\psi_0$ we should first find a solution of
equation
\begin{equation}\label{c:eh1}
H_1 \psi_1 =  E_1 \psi_1\;.
\end{equation}
Looking for it in the form
\begin{equation}
\psi_1 = ( 1 + \beta \xi ) \widetilde\psi\;,
\end{equation}
one finds that $\widetilde\psi$ is a solution of an effective
one-particle Dirac-Coulomb equation
\begin{equation}\label{c:h:tilde}
  \left[ \alphab \cdot {\mathbf p} + \beta \widetilde m - \frac{\widetilde{Z\alpha}}{Z\alpha}
  V_C(r)  \right] \widetilde \psi =
  \widetilde E \widetilde \psi
\end{equation}
with an effective mass
\begin{equation}\label{c:effm}
  \widetilde m = \frac{m\left( 1-\frac{E_1}{M} \right)}{\sqrt{1-\left(\frac{m}{M}\right)^2}}
\end{equation}
and an effective Coulomb coupling constant
\begin{eqnarray}\label{c:effa}
  \widetilde{\Za} &=& \frac{Z\alpha}{\sqrt{1-\left(\frac{m}{M}\right)^2}}\nonumber\\
  &=&
  Z\alpha \left[
    1  + \calO{ \left( \frac{m}{M} \right)^2 }
  \right]\label{c:effa1}
  \,,
\end{eqnarray}
where
\begin{eqnarray}\label{c:xi}
  \xi
  &=&
  \frac{M}{m} \left( 1-\sqrt{1-\left(\frac{m}{M}\right)^2} \right)\nonumber\\
  &=&
  \frac{m}{2M} \left[ 1 + {\cal O} \left( \left( \frac{m}{M} \right)^2 \right) \right]\label{c:xi1}
  \,.
\end{eqnarray}

The solutions of Eq.~(\ref{c:h:tilde}) are similar to the well-known
solutions of the conventional Dirac-Coulomb equation (see, e.g.
\cite{IV}), with the only difference being that the parameters $m$ and
$Z\alpha$ must be replaced by effective values $\widetilde m$ and
$\widetilde{\Za}$, defined in (\ref{c:effm}) and (\ref{c:effa}).

The energies $E_1$ are related to the known eigenvalues of the effective
equation (\ref{c:h:tilde}), $\widetilde E$, by the equation
\begin{equation}\label{c:ee1}
  \widetilde E = \frac{E_1-\frac{m^2}{M}}{\sqrt{1-(\frac{m}{M})^2}}
  \,.
\end{equation}

The eigenvalues and  eigenfunctions of Hamiltonian $H_0$ in
Eq. \eq{Dirac_C}, according to \eq{Dirac2_C}), are related to $E_1$ and
$\psi_1$, as
\begin{eqnarray}
\label{c:e0}
  E_0&=&E_1 + \frac{E_1^2-m^2}{2M}\\
  &=& \widetilde E + \frac{{\widetilde E}^2+m^2}{2M} + \calO{\frac{m^3}{M^2}}
  \;,\\
\label{c:psi0}
  \psi_0
  &=&
  N\, \left[ 1-\frac{1}{4M} [ \mathbf{p}^2, W_C ] + \calO{\left(\frac{m}{M}\right)^2 \Zab^4} \right]\nonumber\\
  &~&\times
  ( 1 + \beta \xi ) \widetilde\psi
  \,,
\end{eqnarray}
where $N$ is a normalization constant, for which one can find (see,
e.g., \cite{decay})
\begin{eqnarray}
  N^2
  &=&
  \frac{1}{1+2\xi \widetilde E/\widetilde m+\xi^2}\\
  &=&
  1-\frac{m}{M} + \frac{\Zab^2}{2n^2} \frac{m}{M}\nonumber\\
  &&+
  {\cal O} \left( \left( \frac{m}{M} \right)^2 \right)
  +
  {\cal O} \left( \frac{m}{M} \Zab^4 \right)
  \,.
\end{eqnarray}

This evaluation is not yet  completed. We note that the energy $E_0$
is expressed in terms of $E_1$ (\ref{c:e0}), and the latter in terms
of $\widetilde{E}$ (\ref{c:ee1}). Meanwhile, $\widetilde{E}$ is a
function of $\widetilde{m}$ (\ref{c:effm}) and $\widetilde{Z\alpha}$
(\ref{c:effa}). The effective mass $\widetilde{m}$ in its turn
depends on $E_1$ as follows from Eq.~(\ref{c:effm}).

To proceed further, we note that for the Dirac-Coulomb problem
\begin{equation}\label{tm:dc}
E_{DC} =  f_{C}(Z\alpha)\,m \,,
\end{equation}
and thus the value of
\begin{equation}\label{tildeF:def}
\widetilde{F} = \frac{\widetilde{E}}{\widetilde{m}}\,,
\end{equation}
being equal to $f_{C}(\widetilde{Z\alpha})$, does not depend on the
effective mass of the orbiting particle $\widetilde{m}$, while the
effective charge $\widetilde{Z\alpha}$, as follows
from Eq.~(\ref{c:effa}), does not depend on energy. This allows
simplifications.

Applying Eqs.~(\ref{tildeF:def}) and (\ref{c:effm}) to
(\ref{c:ee1}), we obtain
\begin{equation}\label{tm:e1f}
E_1= m \frac{\widetilde{F}+\frac{m}{M}}{1+\frac{m}{M}\widetilde{F}}
\,,
\end{equation}
and, using (\ref{c:e0}),
\begin{eqnarray}\label{tm:exa1}
E_0 &=& m +
m\left(1-\frac{m}{M}\right)(\widetilde{F}-1)
\nonumber\\
&& - \frac{m^2}{2M} (\widetilde{F}-1)^2
\frac{\left(1-\frac{m}{M}\right)\left(1+\frac{m}{M}+2\frac{m}{M} \widetilde{F}\right)}
  {\left(1+\frac{m}{M} \widetilde{F}\right)^2}\nonumber\;.
\end{eqnarray}

Since
\[
\widetilde{F}-1={\cal O}\left((Z\alpha)^2\right)\;,
\]
we can efficiently expand
\begin{eqnarray}\label{tm:exp1}
E_0 &=& m +m\left(1-\frac{m}{M}\right)(\widetilde{F}-1) \nonumber\\
&& - \frac{m^2}{2M}(\widetilde{F}-1)^2
\frac{\left(1-\frac{m}{M}\right)\left(1+3\frac{m}{M}\right)}{\left(1+\frac{m}{M}\right)^2}
 \nonumber\\
&& + \calO{m\left(\frac{m}{M}\right)^3\Zab^6}\;,\\
\widetilde m &=& m\sqrt{\frac{1-\frac{m}{M}}{1+\frac{m}{M}}}
  \left[
  1
  -\frac{\frac{m}{M}}{1+\frac{m}{M}} (\widetilde{F}-1)\right.\nonumber\\
&&
  +\frac{\left(\frac{m}{M}\right)^2}{\left(1+\frac{m}{M}\right)^2}
  (\widetilde{F}-1)^2
 \nonumber\\
&&    + \left.\calO{\left(\frac{m}{M}\right)^3\Zab^6}
  \right]\label{tm:exp2}
\,.
\end{eqnarray}

For the pure Coulomb problem it is sufficient to transform
Eq.~(\ref{tm:exp1}), neglecting terms of order $(Z\alpha)^4(m/M)^2m$.
We note, comparing $\widetilde{F}$ and
\[
F=f_{C}({Z\alpha})\;,
\]
that we have to distinguish between $\widetilde{Z\alpha}$ and
${Z\alpha}$ only in the leading term of $(\widetilde{F}-1)$
\[
F= 1 + \left( \frac{{Z\alpha}}{\widetilde{Z\alpha}} \right)^2 ({\widetilde F}-1)
+ {\cal O} \left(  \Zab^4\left(\frac{m}{M}\right)^2m\right)\;.
\]

As a
result, we eventually find for the Coulomb problem
\begin{eqnarray}\label{tm:masc}
E &=& m + m_R (F-1)-\frac{m_R^2}{2M}\left(F-1\right)^2 \,,
\end{eqnarray}
which has corrections only of order $(Z\alpha)^4(m/M)^2m$.

Here we have taken into account that for a pure Coulomb problem
$\delta H=0$ and thus the eigenvalues of the Hamiltonians $H$ in
Eq.~(\ref{Dirac_CH}) and $H_0$ in Eq.~(\ref{c:hoh1}) are the same, i.~e.
$E=E_0$.

This evaluation, following \cite{Gro67}, eventually presents
eigenvalues and eigenfunctions of the Grotch equation
(\ref{Dirac_C}) in terms of the well-known solution of the
Dirac-Coulomb problem (see, e.g., \cite{IV}), but with effective
parameters $\widetilde m$ and $\widetilde{\Za}$. We briefly overview
those solutions in Appendix~\ref{s:dc} (see, e.g., \cite{IV} for
details).

We note that the Grotch equation (\ref{Dirac_C}) and its solution
(\ref{tm:masc}) is a complete account of the static
one-photon-exchange, once we average over the nuclear spin. The
relativistic energies (see, e.g., \cite{IV}) are listed in
Appendix~\ref{s:dc}. We have
not evaluated the wave functions, but it is more appropriate to
perform such an evaluation once we clarify what accuracy is required.
The energy levels
(\ref{tm:masc}) by themselves are obtained without any need for
explicit expressions for the wave functions. However, once we step
out from a pure Coulomb case the wave functions will be required;
however, they are to appear in calculations of a small perturbation and do
not need a high accuracy.

We remind that the energy levels (\ref{c:e0}) and (\ref{tm:masc})
and wave functions (\ref{c:psi0}) obtained above reproduce correctly
\begin{itemize}
\item the leading nonrelativistic term (i.e. a result of the
Schr\"odinger-Coulomb
problem with the reduced mass) exactly in $m/M$;
\item the relativistic corrections (exactly in $Z\alpha)$ for a
infinitely heavy nucleus  (i.e. a result of the Dirac-Coulomb
problem);
\item the leading relativistic recoil correction to energy in order
$(Z\alpha)^4m^2/M$.
\end{itemize}
The result for the energy has to contain also various higher-order
contributions $(Z\alpha)^km^2/M$ ($k\ge6$), which, without being a
complete result, still have a certain sense, since it is sometimes
clear how to upgrade them to a complete result
\cite{shabaev,yelkhovsky}.


\section{Consideration of an arbitrary nonrelativistic-type potential}

Let us consider now a potential, which is a sum of the Coulomb
potential and a ``nonrelativistic-type potential''
\[
V=V_C + V_{N}\;.
\]
The ``nonrelativistic-type potential'' $V_{N}(r)$ is such a potential
that the leading nonrelativistic correction to energy is of order
$\someletter(Z\alpha)^2m$ and the leading relativistic
correction is of order of $\someletter(Z\alpha)^4m$, while the
leading correction to the wave function is of relative order
$\someletter$ both for nonrelativistic and relativistic behavior.
It is understood that $\someletter$ is a small but finite parameter,
such as $\alpha/\pi$, and that the potential $V_{N}(r)$ is smaller than the
Coulomb potential in any area
by a factor of $\someletter$.

We consider such a potential as a nonrelativistic-type potential,
because its relativistic correction, similarly to the case of pure
Coulomb potential, can be found through a relativistic expansion,
which treats relativistic corrections as additional effective terms
of a Hamiltonian of a nonrelativistic Schr\"odinger equation. Such a
consideration is valid, e.g., for the eVP effects in muonic atoms,
but not valid for eVP effects in ordinary atoms.

What is different for consideration of $V_C + V_{N}$ in comparison
with a pure Coulomb problem \cite{Gro67}, reviewed in the previous
section:
\begin{itemize}
\item It is not necessary that $\someletter(Z\alpha)^4m^2/M$
contributions can be
calculated in the one-photon exchange approximation. We suggest
that it is valid
for all $\someletter(Z\alpha)^4m^2/M$ terms, and that sets a
constraint on effects which may be taken into account by the method
developed here. This question is common for Grotch-type and
Breit-type calculations and was discussed for one-loop eVP corrections in
\cite{pra}. As explained there, there is a gauge, where the eVP contribution can be
calculated within such an approximation.
\item Rigorously speaking, there is no such a thing as just
``potential''. One has to deal with a generalized one-photon
exchange. The correction can be due to the photon propagator
correction (as it is in the case of eVP effects), nuclear structure
etc. While its $D_{00}$ component in a static regime is related to a
``potential'' for the external field approximation, the result for the
other terms depends on the nature of the correction. There is no
single rule on how to express the complete effect in terms of
$V_{N}$. Here, we suggest that the expression (\ref{Dirac_C}) holds
for the one-photon contribution in order up to
$\someletter(Z\alpha)^4m^2/M$. The Hamiltonian is of the form
\begin{eqnarray}\label{Dirac_NH}
  H&=&\biggl(
    \alphab \cdot {\mathbf p}
    + \beta m
    + \frac{\mathbf{p}^2}{2M}
    + V
    + \frac{1}{2M} \left\{ \alphab \cdot {\mathbf p}, V
    \right\}\nonumber\\
    &~&~~~ +\frac{1}{4M}
      \left[
         \alphab \cdot {\mathbf p},
         [\mathbf{p}^2, W]
      \right]
  \biggr)\;,
\end{eqnarray}
where
\begin{equation}
W = W_C + W_N\;,
\end{equation}
and an appropriate $W_N$ term is to be found.

Furthermore, we suggest that in general the behavior of $W_N$ is
somewhat similar to that of $\someletter W_C$, and the order of
magnitude of related
matrix elements can be found from that similarity. It is essential
that in some way the last term in the Hamiltonian resulted
from the lower (smaller) component of the nuclear spinor, so the related
matrix elements are of order $(Z\alpha)^4m^2/M$ and may additionally
contain $\someletter$.

All that apparently sets another constraint on interactions
which can be described by means of a Grotch-type equation. What is
important for our purposes is that for the eVP contributions we deal
with a certain correction to the photon propagator, the equation
(\ref{Dirac_C}) is valid and the appropriate function $W_N$ can be
explicitly found (see \cite{pra,II,III}).

\item Next we note that the addition to the Hamiltonian,
defined in Eq.~(\ref{c:dh}),
which vanishes in the pure Coulomb case, does not in
the general situation:
\begin{equation}\label{n:dh}
  \delta H =
  -\left( \frac{V^2}{2M} + \frac{1}{4M} [V,[\mathbf{p}^2,W]]
  \right)\neq 0
      \,.
\end{equation}
In the former, pure Coulomb, case this addition was equal to zero.
It consisted of two operators, matrix elements of which are of order
$(Z\alpha)^4m^2/M$:
\begin{equation}
\bigg\langle \frac{V^2}{2M} \bigg\rangle -\bigg\langle \frac{1}{4M} [V,[\mathbf{p}^2,W]] \bigg\rangle = {\cal O} \left(  \Zab^4\frac{m^2}{M}\right)\;.\nonumber
\end{equation}

These are operators which have a non-vanishing
matrix element between upper-upper (large-large) components of the
muon spinor. To obtain the leading term of order $(Z\alpha)^4m^2/M$
it is sufficient to work with the nonrelativistic wave functions,
$\psi_{\rm NR}$.

So, the equations for the Hamiltonian and the energy are now
\begin{eqnarray}\label{Dirac_N1}
  H &=&  H_0+\delta H \,,\\
\label{Dirac_N2}
  H_0&=&H_1
  + \frac{H_1^2-m^2}{2M} + \frac{1}{4M} [H_1,[\mathbf{p}^2,W]]
  \,,\\
\label{Dirac_N3}
  H_1 &=& \alphab \cdot {\mathbf p} + \beta m + V \frac{1-\beta m/M}{1-(m/M)^2}
  \,,
\end{eqnarray}
where we neglect the terms of order $(Z\alpha)^4(m/M)^2m$, and
\begin{eqnarray}\label{Dirac_N4}
  E &=&  E_0+\delta E\nonumber\\
  \delta E&=&\langle \psi_{\rm NR} \vert \delta H  \vert \psi_{\rm NR} \rangle
  \,.
\end{eqnarray}
Since $\delta E$ is already of order $\someletter(Z\alpha)^4m^2/M$,
only the linear corrections are necessary and the nonrelativistic
wave function is that of the problem with $H_0$.

In the first order in $\someletter$ we need only pure Coulomb wave
functions (see Appendix~\ref{s:dc}), since we explicitly took into
account that $\delta H$, which vanishes in the pure Coulomb case,
has to be proportional to $\someletter$. To second order in
$\someletter$ we have to construct the nonrelativistic wave function
perturbatively. Such a problem can be successfully resolved for many
problems numerically.
\item The solution suggests that the effective energy $\widetilde{E}$
depends on the effective mass $\widetilde{m}$, and the actual energy
$E_0$ is expressed in terms of $\widetilde{E}$. Meantime, the
effective mass $\widetilde{m}$ depends on the energy $E_0$. In the
case of the pure Coulomb problem, the ratio
\[
\frac{\widetilde{E}}{\widetilde{m}}=\widetilde{F}
\]
does not depend on the effective mass and as a result we can
disentangle $\widetilde{E}$ and $\widetilde{m}$. In general case,
the ratio $\widetilde{E}/\widetilde{m}$ depends on $\widetilde{m}$
and, through it, it depends on energy $E_0$. This can be resolved
only through expansion over the relativistic effects.

We have to apply expressions~(\ref{tm:exa1}) and~(\ref{tm:exp2})
studied above, where now the solution of the Dirac equation with
potential $V$ is of the form
\begin{equation}\label{tm:n:dc}
E =  f_{D}(Z\alpha, Z\alpha m/\mu)\,m \,,
\end{equation}
and
\[
\widetilde{F}=f_{D}(\widetilde{Z\alpha},\widetilde{Z\alpha}\widetilde{m}/\mu)\;
\]
where $f_{D}$ is a dimensionless energy of the Dirac equation with $V$
and
\[
\widetilde{F}-1={\cal O}\left((Z\alpha)^2\right)\;.
\]
In contrast to the pure Coulomb case the dimensionless energy $f_{D}$
depends on the effective mass through a dimensionless parameter
$\widetilde{Z\alpha}\widetilde{m}/\mu$. This is possible if the
potential $V_N$ depends on the dimensional parameter $\mu$. While
calculating various integrals over the wave function the scale
parameter of the potential, say, ``radius'' ($\sim 1/\mu$), is
naturally compared with the atomic Bohr radius ($\sim 1/Z\alpha m$).
For instance, in the case of eVP corrections in muonic atoms $\mu=m_e$
and the related parameter is $\sim 1.5 Z$.

Next we note (see Eq.~(\ref{tm:exp2})) that
\[
\widetilde{m} = m_0\left(1 + {\cal O}\left(\frac{m}{M}(\widetilde{F}-1)\right) +
\dots\right)
\]
where $m_0$ is the result in the limit $Z\alpha\to0$. The
relativistic part is already proportional to $(Z\alpha)^4m$ and it
is sufficient to apply $m_0$ there. The nonrelativistic part is of
order $(Z\alpha)^2m$ and a correction of relative order
$(Z\alpha)^2m/M$ is important in the leading approximation, while
higher powers of $m/M$ are to be neglected here.

The result of the expansion with all terms required is
\begin{eqnarray}
\widetilde m &=& m_R\sqrt{1-\left(\frac{m}{M}\right)^2} \left[
  1
  -\frac{m}{M}(\widetilde{F}-1)\right]\nonumber\\
&=&  m_R\sqrt{1-\left(\frac{m}{M}\right)^2} -\frac{m}{M} E_{\rm NR}
  \label{tm:n:exp2}
\,,
\end{eqnarray}
where $E_{\rm NR}$ is the nonrelativistic part of the energy for
the Schr\"odinger problem with $V$. As we mentioned, any further
$m/M$ corrections in the second term are unimportant and in
particular, we can choose to calculate $E_{\rm NR}$ with a muon mass
$m$ or with the reduced mass $m_R$.

The effective mass is not included in $\widetilde F$ and $F$ directly,
but only in a combination
\begin{eqnarray}
\widetilde{Z\alpha} \widetilde m &=& Z\alpha m_R\left[1-\frac{E_{\rm
NR}}{M} \right]
  \label{tm:n:exp3}
\,.
\end{eqnarray}

Thus we find
\begin{eqnarray}\label{tm:n:F}
\widetilde{F}-1&=&\frac{(Z\alpha)^2}{(\widetilde{Z\alpha})^2}
(f_{D}(\widetilde{Z\alpha},\widetilde{Z\alpha}\widetilde{m}/\mu)-1)\nonumber\\
&=&
\frac{(Z\alpha)^2}{(\widetilde{Z\alpha})^2}
\Biggl\{
f_{D}\left(\widetilde{Z\alpha},Z\alpha m_R/\mu\right)-1\nonumber\\
&&-\frac{E_{\rm NR}}{M}\,\kappa\frac{\partial}{\partial \kappa}
f_{D}\left(\widetilde{Z\alpha},\kappa\right)
\Biggr\}
\;,
\end{eqnarray}
where for the following it is useful to introduce
\[
  \kappa=\frac{Z\alpha m_R}{\mu}
  \;.
\]

One can treat the first two terms in (\ref{tm:n:F}) separately, introducing
\[
F_0-1=
\frac{(Z\alpha)^2}{(\widetilde{Z\alpha})^2}
\left\{
f_{D}\left(\widetilde{Z\alpha},Z\alpha m_R/\mu\right)-1
\right\}
\;,
\]
which now does not depend on $\widetilde m$. The energy can also be split
into two terms
\[
E_0=E^{(1)}+E^{(2)}\;,
\]
with the first term similar to the one for the pure Coulomb case
(cf. Eq.~\ref{tm:masc})
\begin{eqnarray}\label{tm:n:masc1}
E^{(1)} &=& m + m_R (F_0-1)-\frac{m_R^2}{2M}\left(F_0-1\right)^2\;.
\end{eqnarray}
For the second term we note that
$(f_D-1)$ is the leading nonrelativistic contribution to the energy
and with a sufficient accuracy we can approximate
\[
f_D\left(\widetilde{Z\alpha},Z\alpha m_R/\mu\right)-1 =
\frac{E_{\rm NR}}{m_R}
\]
and thus
\begin{eqnarray}\label{tm:n:masc2}
E^{(2)} &=& -\frac{m_R^2}{2M}\frac{\partial}{\partial
\ln\kappa}\left(\frac{E_{\rm NR}}{m_R}\right)^2\,.
\end{eqnarray}

Eventually we arrive at the identity for the complete energy
\begin{eqnarray}\label{tm:n:masc}
E &=& m + m_R (F_0-1)-\frac{m_R^2}{2M}\left(F_0-1\right)^2\nonumber\\
&-&\frac{m_R^2}{2M}\frac{\partial}{\partial
\ln\kappa}\left(\frac{E_{\rm NR}}{m_R}\right)^2\nonumber\\
&-&\langle \psi_{\rm NR} \vert  \left( \frac{V^2}{2M} + \frac{1}{4M}
[V,[\mathbf{p}^2,W]]
  \right) \vert \psi_{\rm NR} \rangle\,,
\end{eqnarray}
which is valid for our purposes and have corrections
to order $(Z\alpha)^4m^3/M^2$ and $\someletter(Z\alpha)^4m^3/M^2$.

For the relativistic recoil term $(Z\alpha)^4m^2/M$ we choose
between applying $m_R$ and $m$ in such a way that it would simplify
a comparison with Breit-type calculations of the same corrections
(see \cite{pra}) for details. A difference between $m_R$ and $m$ in
relativistic recoil corrections produces only terms of order $(Z\alpha)^4(m/M)^2m$.

\item In contrast to the pure Coulomb problem in the external field
approximation, for which we know the
energy and wave functions in closed analytic form, we indeed cannot
know them for an arbitrary potential.

For the main term in (\ref{tm:n:masc}) we need to be able to find
the energy of the Dirac equation with potential $V$ and the reduced mass
$m_R$ with a required accuracy. For two other terms we need to know
only the nonrelativistic results for the related problem of a
Schr\"odinger equation with potential $V$ and the reduced mass
$m_R$.

Both relativistic and nonrelativistic problems can be considered at
this stage perturbatively since\linebreak $\someletter\ll1$ and $V_N$ is a
small correction to $V_C$.
\end{itemize}

\section{Conclusions}

The main result of this paper is that there is a certain kind of
potential $V$ for which a calculation of the relativistic effects
can be split into two parts. One is a calculation of the energy in
the external field approximation for a muon with the mass equal to
its reduced mass in the atom. That is a ``standard'' problem of a
Dirac equation for a particle with the mass equal to the reduced
mass. This calculation can be, in principle, performed by various
means, including numerical solutions.

The second part, which is a non-trivial part of the relativistic
recoil correction, can be obtained once we know the nonrelativistic
results for the atom with a muon with the reduced mass. That
includes certain derivatives. Such a reduction of the relativistic
correction to nonrelativistic calculations essentially simplifies the
problem. Roughly speaking, the essential two-body effects are less
complicated than the one-particle relativistic problem.

Apparently, a number of problems to be solved for a relativistic muon
is limited and we do not expect that a Dirac equation with potential
$V$ can be solved exactly. As far as the non-Coulomb term is a
perturbation, i.e. for $\someletter\ll1$, we can find all required
elements perturbatively.

In particular, in the subsequent paper \cite{II} we apply the
developed approach to the eVP corrections in the first order in
$\alpha$, i.e. to the relativistic Uehling correction. In this case,
one can expand (\ref{tm:n:masc}) in $\someletter=\alpha/\pi$ and find
that all required terms are known in a closed form. In the other
subsequent paper \cite{III} the same master equation is applied to
the relativistic recoil K\"allen-Sabry correction,
however, none of the eVP related terms are known analytically. So,
they are calculated by means of numerical integration. Here, it is
still sufficient to work in the first order in
$\someletter=(\alpha/\pi)^2$. However, the relativistic recoil results
of the same order,
namely $\alpha^2(Z\alpha)^4m^2/M$ arise also from double iteration
of the Uehling potential, for this case $\someletter=\alpha/\pi$,
and the second order in $\someletter$ terms are required in (\ref{tm:n:masc}).
The recoil effects are obtained for these corrections
also by means of numerical integration \cite{III}.

To conclude, we mention that the condition $\someletter\ll1$ was set
only because we are interested in developing a framework for perturbative
calculations of the eVP
relativistic recoil effects, which are performed in subsequent papers
\cite{II,III}. In principle, one can consider any
``nonrelativistic-type potential'', but the related Dirac equation
should be solved numerically.

\section*{Acknowledgments}

This work was supported in part by DFG under grant GZ: HA 1457/7-2
and RFBR under grant No.~12-02-31741. A part of the work was done
during a stay of VGI and EYK at  the Max-Planck-Institut f\"ur
Quantenoptik, and they are grateful to it for its  warm hospitality.

\appendix

\section{Solution of the Dirac equation with Coulomb potential\label{s:dc}}


The exact relativistic energy for  a pure Dirac-Coulomb problem
$E_C(nl_j)$ for the $nl_j$ state is of the form (see, e.g.,
\cite{IV})
\begin{eqnarray}\label{ECnlj}
  E_C(nl_j) &=&  f_C(Z\alpha)\,m\\
  f_C(Z\alpha)&=& \frac{1}{\sqrt{1+\frac{\Zab^2}{(n_r+\zeta)^2}}}
\end{eqnarray}
and\footnote{It is customary to use $\kappa$ for $(-1)^{j+l+1/2}
(j+1/2)$ (cf. \cite{IV}), however, $\kappa$ is used in our papers on
muonic atoms for something else.}
\begin{eqnarray}
\nu &=& (-1)^{j+l+1/2} (j+1/2) \,,\nonumber\\
\zeta &=& \sqrt{\nu^2 -(Z\alpha)^2} \,,\nonumber\\
 n_r &=& n - |\nu| \,.\nonumber
\end{eqnarray}


The wave functions of the  Dirac-Coulomb problem are (see, e.g.,
\cite{IV})
\begin{eqnarray}
 \psi_{njlm}^{(C)}({\bf r})
 &=&
 \left(
 \begin{array}{c}
   \Omega_{j,l,m}({\bf r}/r) \; f(r) \\
   (-1)^{\frac{1+2l-2j}{2}} \Omega_{j,2j-l,m}({\bf r}/r)
   \; g(r)
 \end{array}
 \right)
 \,,\\
 &&\nonumber
\end{eqnarray}
where the radial components are
\begin{eqnarray}
 \left.\begin{array}{c}f\\g\end{array}\right\}
 &=&
 \pm \frac{(2m\eta)^{3/2}}{\Gamma(2\zeta+1)}
 \sqrt{
  \frac{(m\pm E_C)\Gamma(2\zeta+n_r+1)}
  {\frac{4\Za m}{\eta} \left( \frac{\Za}{\eta} - \nu \right)
   n_r!}
 }
 \nonumber\\
 &\times&
 e^{-m\eta r}
 (2m\eta r)^{\zeta-1}
 \nonumber\\
 &\times&
 \Biggl\{
 \left(
  \frac{\Za}{\eta} - \nu \right)
  {}_1F_1(-n_r,2\zeta+1\,;2m\eta r)
  \nonumber\\
  &&\mp\;
  n_r \times {}_1F_1(1-n_r,2\zeta+1\,;2m\eta r)
 \Biggr\}\;.
\end{eqnarray}
Here the upper signs correspond to the large component $f$ and lower
ones are for the small components $g$; ${}_1F_1(a,b;z)$ are
confluent hypergeometric functions, $\Omega_{jlm}$ is a
spherical spinor and
\begin{eqnarray}
\eta &=& \sqrt{1 - (E_{nl_j}/m)^2} \nonumber\\
&=&\frac{Z\alpha}{\sqrt{(n_r+\zeta)^2+\Zab^2}} \,.\nonumber
\end{eqnarray}

The leading nonrelativistic contribution to the Dirac-Coulomb wave
functions can be expressed in terms of the eigen functions of the
Schr\"odinger-Coulomb problem
\begin{equation}
  \Phi_{nlm}^{(C)}({\bf r}) = Y_{lm}({\bf r}/r) R_{nl}(r)
  \;,
\end{equation}
where
\begin{eqnarray}
  R_{nl}(r)
  &=&
  \frac{2(\Za m)^{3/2}}{n^{l+2}(2l+1)!}
  \,
  \sqrt{\frac{(n+l)!}{(n-l-1)!}}
  \nonumber\\
  &\times&
  (2\Za m r)^l
  \,
  e^{-\frac{\Za m r}{n}}
  \nonumber\\
  &\times&
  {}_1F_1\left(
  -n+l+1,2l+2; \frac{2\Za m r}{n}
  \right)
\end{eqnarray}
and $Y_{lm}$ are spherical functions.


\newpage

\begin{thebibliography}{19.}

\frenchspacing

\bibitem{nature}
R. Pohl, A. Antognini, F. Nez, Fernando D. Amaro, F. Biraben, J. M.
R. Cardoso, D. S. Covita, A. Dax, S. Dhawan, L.M.P. Fernandes, A.
Giesen, T. Graf, T.W. H\"ansch, P. Indelicato, L. Julien, Cheng-Yang
Kao, P. Knowles, E.-O. Le Bigot, Yi-Wei Liu, J.A.M. Lopes, L.
Ludhova, C.M.B. Monteiro, F. Mulhauser, T. Nebel, P. Rabinowitz,
J.M.F. dos Santos, L.A. Schaller, K. Schuhmann, C. Schwob, D. Taqqu,
J.F.C.A. Veloso and F. Kottmann, Nature (London) {\bf 466}, 213
(2010);\\
A. Antognini, F. Nez, K. Schuhmann, F.D. Amaro, F. Biraben, J.M.R.
Cardoso, D.S. Covita, A. Dax, S. Dhawan, M. Diepold, L.M.P.
Fernandes, A. Giesen, A.L. Gouvea, T. Graf, T.W. H\"ansch, P.
Indelicato, L. Julien, C.-Y. Kao, P. Knowles, F. Kottmann, E.-O. Le
Bigot, Y.-W. Liu, J.A.M. Lopes, L. Ludhova, C.M.B. Monteiro, F.
Mulhauser, T. Nebel, P. Rabinowitz, J.M.F. dos Santos, L.A.
Schaller, C.Schwob, D.Taqqu, J.F.C.A. Veloso, J. Vogelsang, and R.
Pohl, Science, {\bf 339}, 417 (2013).

\bibitem{codata2010} P.J. Mohr, B.N. Taylor, and D.B. Newell, Rev. Mod. Phys. {\bf 84}, 1527 (2012).


\bibitem{mainz}
J.C. Bernauer, P. Achenbach, C. Ayerbe Gayoso, R. B\"ohm, D. Bosnar,
L. Debenjak, M.O. Distler, L. Doria, A. Esser, H. Fonvieille, J.M.
Friedrich, J. Friedrich, M. G\'omez Rodríguez de la Paz, M. Makek,
H. Merkel, D.G. Middleton, U. M\"uller, L. Nungesser, J.
Pochodzalla, M. Potokar, S. S\'nchez Majos, B.S. Schlimme, S.
\v{S}irca, Th. Walcher, and M. Weinriefer, Phys. Rev. Lett. {\bf
105}, 242001 (2010);\\ Phys. Rev. Lett. {\bf 107}, 119102 (2011).

\bibitem{zumbro}
J.D. Zumbro, E.B. Shera, Y. Tanaka, C.E. Bemis, Jr., R.A. Naumann,
M.V. Hoehn, W. Reuter, and R.M. Steffen, Phys. Rev. Lett. {\bf 53},
1888 (1984).


\bibitem{mumass} I. Beltrami, B. Aas, W. Beer, G.
De Chambrier, P.F.A. Goudsmit, T. v. Ledebur, H.J. Leisi, W.
Ruckstuhl, W.W. Sapp, G. Strassner, and A. Vacchi, Nucl. Phys. A{\bf
451}, 679 (1986).

\bibitem{pimass}
N. Nelms, D.F. Anagnostopoulos, M. Augsburger, G. Borchert, D.
Chatellard, M. Daum, J.-P. Egger, D. Gotta, P. Hauser, P.
Indelicato, E. Jeannet, K. Kirch, O.W.B. Schult, T. Siems, L.M.
Simons, and A. Wells, NIM A{\bf 477}, 461 (2002);\\
D. Gotta, Prog. Part. Nucl. Phys. {\bf 52}, 133 (2004).

\bibitem{mep2:1}
M. Hori, A. Dax, J. Eades, K. Gomikawa, R.S. Hayano, N. Ono, W.
Pirkl, E. Widmann, H.A. Torii, B. Juh\'asz, D. Barna, and D.
Horv\'ath, Phys. Rev. Lett. {\bf 96}, 243401 (2006).

\bibitem{mep2:2}
M. Hori, A. S\'ot\'er, D. Barna, A. Dax, R. Hayano, S. Friedreich,
B. Juh\'asz, T. Pask, E. Widmann, D. Horv\'ath, L. Venturelli, and
N. Zurlo, Nature {\bf 475}, 484 (2011).


\bibitem{shabaev} V.M. Shabaev, Theor. Math. Phys., {\bf 63} 588
(1985).

\bibitem{yelkhovsky} A.S. Yelkhovsky, arXiv:hep-th/9403095v2.

\bibitem{Gro67}
H.~Grotch and D.R.~Yennie, Z.~Phys.,  {\bf 202}, 425 (1967).

\bibitem{jentschura} U. D. Jentschura, Phys. Rev. A {\bf 84}, 012505
(2011).

\bibitem{borie11}
E. Borie, Annals of Physics {\bf 327}, 733 (2012).

\bibitem{pra} S.G. Karshenboim, V.G. Ivanov, and E.Yu. Korzinin, Phys. Rev. A {\bf 85},
032509 (2012).


\bibitem{pach96}
K.~Pachucki, Phys. Rev.A, \textbf{53}, 2092, (1996).

\bibitem{pach04}
A.~Veitia and K.~Pachucki, Phys. Rev. A. \textbf{69}, 042501 (2004).

\bibitem{borie05}
E. Borie, Phys. Rev. A \textbf{71}, 032508 (2005)

\bibitem{a2Za4} E.Yu. Korzinin, V.G. Ivanov, and S.G. Karshenboim, Phys. Rev. D,
{\bf 88}, 125019 (2013); arXiv:1311.5784.

\bibitem{II} V.G. Ivanov, E.Yu. Korzinin, and S.G. Karshenboim
Phys. Rev. A \textbf{89}, 022103 (2014); arXiv:1311.5790.

\bibitem{III} S.G. Karshenboim, E.Yu. Korzinin, and V.G. Ivanov, Phys. Rev. A,
{\bf 89}, 032129 (2014); arXiv:1311.5822.

\bibitem{uehl_cl} A.B. Mickelwait and H.C. Corben, Phys. Rev. {\bf 96}, 1145 (1954);\\
G.E.~Pustovalov, Zh. Eksp. Teor. Fiz. {\bf 32}, 1519 (1957) [in
Russian];
Sov. Phys. JETP {\bf 5}, 1234 (1957);\\
D.D.~Ivanenko and G.E.~Pustovalov. Usp. Fiz. Nauk {\bf 61}, 27
(1957) [in Russian]; Adv. Phys. Sci. {\bf 61}, 1943 (1957).

\bibitem{uehl_an} S.G.~Karshenboim, V.G.~Ivanov, and E.Yu.~Korzinin,
Eur.~Phys.~J.  D {\bf 39}, 351 (2006).

\bibitem{rel1} S.G.~Karshenboim, Can. J. Phys. {\bf 76}, 169 (1998);\\
ZhETF {\bf 116}, 1575 (1999) [in Russian]; JETP {\bf 89}, 850
(1999).

\bibitem{rel2} E.Yu.~Korzinin, V.G.~Ivanov, and  S.G.~Karshenboim,
Eur.~Phys.~J. D {\bf 41}, 1 (2007).

\bibitem{spin0} D. A. Owen, Phys. Rev. D {\bf 42}, 3534 (1990); D {\bf 46}, 4782
(E) (1992);\\ D. A. Owen, Found. Phys. {\bf 24}, 273 (1994);\\
M. Halpert and D. A. Owen, J. Phys. G {\bf 20}, 51 (1994).

\bibitem{spin1} K. Pachucki and S. G. Karshenboim, J. Phys. B {\bf 28}, L221 (1995).

\bibitem{spin_nov} I. B. Khriplovich, A. I. Milstein, and R. A.
Sen'kov, Phys. Lett. A {\bf 221}, 370 (1996);\\
JETP {\bf 84}, 1054 (1997).

\bibitem{spin_r} J.~L.~Friar, J.~Martorel, D.~W.~L.~Sprung, Phys. Rev. A{\bf 56}, 6, 4579 (1997).


\bibitem{ede} G.P. Lepage. Phys. Rev. A {\bf 16}, 863 (1977);\\
G.T. Bodwin, D.R. Yennie and M.A. Gregorio. Rev. Mod. Phys. {\bf
57}, 723 (1985);\\
J. Sapirstein and D.R. Yennie, in T. Kinoshita (Ed.), {\em Quantum
Electrodynamics\/} (World Sci., Singapore, 1990), p. 560;\\ M.I.
Eides, S.G. Karshenboim and V.A. Shelyuto, Ann. Physics {\bf 205},
231 (1991).

\bibitem{IV}
L.D. Landau and E.M. Lifshitz. \textit{Course of Theoretical
Physics\/}. Vol. 4: V.B. Berestetskii, E.M. Lifshitz and L.P.
Pitaevskii. \textit{Quantum Electrodynamics\/}. (Pergamon Press,
Oxford, 1982).

\bibitem{decay} V. G. Ivanov and S. G. Karshenboim,  Phys. Rev. A \textbf{79},
032518 (2009).

\end{thebibliography}
\end{document}